\documentclass[11pt,a4paper]{article}
\usepackage{jcappub}                    
\usepackage{amsmath,amssymb,mathtools}  
\usepackage{soul}                       
\usepackage[usenames,dvipsnames]{xcolor}

\newcommand{\dd}{\text{d}}
\newcommand{\dt}{\text{d}t}
\newcommand{\dx}{\text{d}x}

\setstcolor{red}    
\setulcolor{red}    
\allowdisplaybreaks 

\title{Adiabatic regularization of power spectra in nonminimally coupled chaotic inflation}

\author{Allan L. Alinea} 

\affiliation{Department of Physics, Osaka University, Toyonaka, Osaka 560-0043, Japan}

\emailAdd{alinea@het.phys.sci.osaka-u.ac.jp}
 
\abstract{We investigate the effect of adiabatic regularization on both the tensor- and scalar-perturbation power spectra in \textit{nonminimally} coupled chaotic inflation. Similar to that of the \textit{minimally} coupled general single-field inflation, we find that the subtraction term is suppressed by an exponentially decaying factor involving the number of $ e $-folds. By following the subtraction term long enough beyond horizon crossing, the regularized power spectrum tends to the ``bare'' power spectrum. This study justifies the use of the unregularized (``bare'') power spectrum in standard calculations.}


\begin{document}
\maketitle

\bigskip
\bigskip
\section{Introduction}
\label{intro}
Cosmic inflation \cite{Guth:1980zm, Starobinsky:1980te, Starobinsky:1979ty, Sato:1980yn, Linde:1983gd, Albrecht:1982wi} found its place in modern cosmology as a solution to the horizon and flatness problems and an explanation for the origin of primordial cosmological perturbations \cite{LiddlenLyth, Mukhanov} that served as seeds for the large-scale structures (e.g., galaxies and clusters of galaxies) nowadays constituting the Universe. Like any other system of ideas intended to explain nature or a part thereof, this theory has to satisfy at least the two-pronged requirement of science namely, self consistency and agreement with experiment. The latter requires a theory to explain existing measurement results before its inception and predict values of physical observables (e.g., in the case of inflation, non-Gaussianity, tensor-to-scalar ratio, spectral tilt, etc. \cite{Planck:2013jfk,Ade:2015lrj}). On the other hand, self-consistency ensures that pathways of ideas within the framework of the theory do not lead to illogical contradiction and problematic calculation results such as those involving non-removable divergences. On this side, inflation must pass checks involving loop corrections \cite{Senatore:2009cf,Seery:2007we}, divergences in predicted physical observables \cite{Parker:Toms}, and the more general question of renormalization (e.g., see Ref. \cite{Contillo:2011ag}).

\begin{figure}[hbt]
	\centering
	\makebox[\textwidth][c]{\includegraphics[scale=0.9]{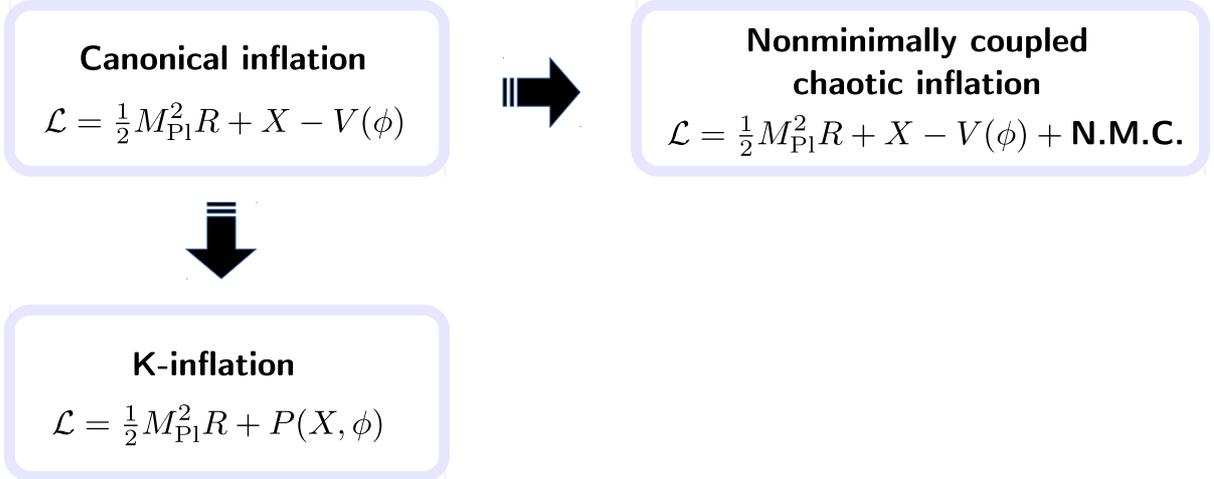}}	
	\caption{Generalizations of the canonical inflation. The canonical case involves the Lagrangian $ \mathcal L = \frac{1}{2}M_\text{Pl}^2 R + X - V(\phi) $, where $ M_\text{Pl},\, R,\, X, $ and $ V $ are the Planck mass, Ricci scalar, kinetic term, and potential term respectively. On the one hand, nonminimally coupled chaotic inflation (right hand most box) involves the addition of nonminimal coupling (N.M.C.) term to the canonical case. On the other hand, $ k $-inflation involves a Lagrangian that is a general functional of $ X $ and $ \phi $; that is, $ P(X, \phi ) $.}
	\label{figFlowOfGen}
	\bigskip
\end{figure}

In this work, we focus on the side of self-consistency and investigate the effect of removing infinities from the two-point functions of scalar and tensor perturbations, on their power spectra. In particular, we consider the adiabatic regularization \cite{Parker:Toms, Parker:1974qw, Fulling:1974pu, Bunch:1980vc} of the power spectra for the perturbations in \textit{nonminimally coupled} single-field inflation \cite{Fakir:1990eg,Futamase:1987ua,Makino:1991sg,Komatsu:1999mt,Nozari:2010uu} (see \textbf{Sec. \ref{setUp}} for the details of the setup). It is an extension/generalization of the work of Urakawa and Starobinsky \cite{Urakawa:2009} involving the adiabatic regularization of the power spectra in the \textit{canonical} single-field inflation. In our previous work \cite{Alinea:2015pza}, we performed the extension/generalization of their study along a different pathway including cases where the speed of sound $ c_s^2, $ is not constant (that is, \textit{general} single-field inflation, or, as usually referred to in the literature, \textit{k}-inflation,) but maintained the \textit{minimal-coupling} aspect of the theory (see \textbf{Fig}. \ref{figFlowOfGen}). This time, adding a nonminimal coupling term on top of the canonical case ($ c_s^2 = 1 $), we wish to determine if adiabatic regularization will significantly modify the power spectrum contrary to what was found by Urakawa and Starobinsky and in our extension to the general single-field inflation of their study.

Before we proceed with the calculation to fulfill this aim, it is good to shed some light on adiabatic regularization in the context of the problem we are dealing in this work. The two-point function of the gauge-invariant scalar perturbation $ \mathcal R $ \cite{Bardeen:1980kt} can be written as
\begin{align}
	\label{twoPtFn}
	\langle \mathcal R(\tau, \mathbf{x})\mathcal R(\tau, \mathbf{y})\rangle
	&=
	\int\frac{\dd k}{k}
	\frac{\sin(k|\mathbf x - \mathbf y|)}{k|\mathbf x - \mathbf y|}
	\Delta^2_{\mathcal R}(k,\tau),
\end{align}
where $ \tau $ is the conformal time, $ (\mathbf x, \mathbf y) $ are pairs of position vectors, $ k $ is the wavenumber, and $ \Delta^2_{\mathcal R} $ is the dimensionless power spectrum given by 
\begin{align}
	\Delta _{\mathcal R}^2(k,\tau )
	=
	\frac{k^3}{2\pi ^2}\big|\mathcal R_k(\tau )\big|^2,
\end{align}
with the subscript $ k $ in $ \mathcal R_k $ indicating momentum space. (Similar expression holds for the tensor perturbation.) The adiabatic condition \cite{Parker:Toms,BirrellDavies} tells us that $ |\mathcal R_k(\tau )| $ scales as $ 1/k $ for large $ k $ leading to $ \Delta^2_{\mathcal R} \sim k^2 $. It follows that in the coincidence limit, $\mathbf x\,\rightarrow \,\mathbf y$, the integral diverges. As applied to this scenario, adiabatic regularization is a method of systematically subtracting out terms in the integrand to yield a finite two-point function, subject to the assumption that the adiabatic condition holds. Such a subtraction scheme is correspondingly reflected in the power spectrum. One may then see that the physical power spectrum is the difference between the ``bare'' and the corresponding subtraction term. 

There is a disagreement about this viewpoint. Parker and his collaborators \cite{Parker:2007,Agullo:2009,Agullo:2009:II} advocate the need for adiabatic regularization and claim that it could significantly modify the power spectrum (with both the ``bare'' and the subtraction term evaluated before or at the moment of horizon crossing), producing expressions not in accord with the ``bare'' power spectrum used in standard calculations. Some other people casted some doubt on the necessity of this subtraction scheme based on the claims that (a) the regularised power spectrum can become negative when one goes beyond the minimal subtraction scheme \cite{Finelli:2007fr} and (b) in the far infrared regime, the adiabatic expansion is no longer valid \cite{Durrer:2009ii}. Moreover, in the review paper of Bastero-Gil et al. \cite{Bastero-Gil:2013}, the authors argued that the cosmic microwave background (CMB) anisotropy variable $ C_l $ depends on the inflaton two-point function evaluated at $ \mathbf x \ne \mathbf y$; thus, practically avoiding the divergence in (\ref{twoPtFn}).

We adopt a position that adiabatic regularization may be necessary and its application to calculation of physical observables produces physically consistent result. In so taking this position, we wish not to devote this paper as an argumentative defense of such position. What we do here is leave some words in favor of our chosen side and focus on the main subject of this work; that is, the \textit{application} of adiabatic regularization to the calculation of power spectra. First, the coincidence limit $ \mathbf x \,\rightarrow \,\mathbf y $ is a mathematical possibility that cannot be avoided by the mere existence of physical circumstances where such a limit may not hold. Furthermore, when one considers potentials involving self interaction or the energy-momentum tensor\footnote{We do not deal with loop corrections, higher-order corrections to the power spectrum, and the energy-momentum tensor. They deserve separate studies of their own.}, the divergences are certainly inevitable. Second, the adiabatic regularization \textit{includes} as its prescription that \textit{the subtraction scheme be minimal}; rather informally, those terms producing infinity should be the only subjects of regularization to remove them. Since the adiabatic expansion say, of $ \mathcal R $, is in general, not convergent but only asymptotic \cite{Parker:Toms}, such a prescription prevents a given quantity---one whose mathematical sickness we want to cure in the first place---from becoming unphysical or ill-behaved (e.g., power spectrum becoming negative). Third and last, determining the definite cut on the value of the spectrum of $ k $ as to when adiabatic regularization should be or should not be performed would seem to introduce an unnecessary complication to this method whenever quantities in momentum space are considered. If only to tame the doubts casted by Durrer et al. \cite{Durrer:2009ii} that the adiabatic expansion may not be valid in the far infrared limit, we note that there are good indications that the subtraction term becomes negligible for the quantities involving  superhorizon modes (see for instance, Refs. \cite{Urakawa:2009,Alinea:2015pza}).

In this work as in our previous work \cite{Alinea:2015pza}, we follow the track laid down by Urakawa and Starobinsky \cite{Urakawa:2009}. We perform adiabatic regularization using the minimal subtraction scheme and follow the subtraction term long after the first horizon crossing during inflation. As what they found out, the subtraction term exponentially decays with the number of $ e $-folds, resulting to the standard expression; i.e., the regularized power spectrum tends to the ``bare'' power spectrum. As already mentioned above, we have done the same for the extension to the minimally coupled general single-field inflation. This study showed that apart from cases that may significantly deviate from the condition of scale invariance, the subtraction ``leads to no difference in the power spectrum'' \cite{Alinea:2015pza}. It is then our aim in this investigation to see if the same holds for the case of nonminimally coupled inflation.

This paper is organized as follows. In \textbf{Sec. \ref{setUp}}, we provide the setup for nonminimally coupled chaotic inflation. In so doing, we state the action and the background equations necessary for our analysis and calculations. In \textbf{Sec. \ref{``bare''Power}}, perturbations are introduced about the FLRW metric and the ``bare'' power spectra for these perturbations (scalar and tensor) are calculated up to first order in the slow-roll parameters. After this, we compute the subtraction terms (also up to first order in the slow-roll parameters) and perform adiabatic regularization of the power spectra in \textbf{Sec. \ref{adiabaticReg}}. Following this, we justify the use of both the Einstein and Jordan frames in our calculation in \textbf{Sec. \ref{remarksEinsteinFrame}}. In the last section, we state our conclusion and future prospects about further generalizing this work to cases involving non-constant speed of sound, higher-order corrections to the power spectrum, and gravity beyond general relativity.

\bigskip
\section{Set up: Nonminimal Chaotic Inflation}
\label{setUp}
Nonminimal chaotic inflation was proposed by Fakir and Unruh \cite{Fakir:1990eg} (see also, Ref. \cite{Futamase:1987ua}) to avoid the fine-tuning problem associated with the coupling constant $ \lambda $ in chaotic inflation involving the potential $ \lambda \phi ^4, $ where $ \phi $ is the inflaton field. Since its conception in 1988, several other papers followed covering (a) a rigorous evaluation of the density perturbation amplitude (1991) \cite{Makino:1991sg}, (b) constraints on physical observables such as the tensor-to-scalar ratio and spectral tilt (1999) \cite{Komatsu:1999mt}, (c) loop corrections (2010, 2015) \cite{Hertzberg:2010dc,Inagaki:2015fva}, (d) observational consequences with the addition of a quadratic potential term (2011) \cite{Linde:2011nh}, and (e) non-Gaussianity (2011) \cite{Qiu:2011}, among others. It is thus, an intensively studied model consistent with the current observational constraints \cite{Planck:2013jfk,Ade:2015lrj}.

We consider in this work the action given by
\begin{align}
	\label{lagrangianNonminimal}
	S
	&=
	\frac{1}{2}\int \sqrt{-g}\,\dd ^4 x
	\Big[
		M_\text{Pl}^2f(\phi )R 
		- 
		g^{\mu \nu }\nabla _\mu \phi \nabla _\nu \phi 
		-
		2V(\phi )
	\Big],
\end{align}
where $ g_{\mu \nu } $ is the metric, $ R $ is the Ricci scalar, $ V(\phi ) $ is the potential, and $ f(\phi ) \equiv 1 + h(\phi ) $ with $ M_\text{Pl}^2 $ as the Planck mass (squared) that from hereon, take as unity for brevity. Furthermore, given $ \xi $ as the nonminimal coupling parameter, we take\footnote{From a quantum field theoretic point of view, the nonminimal coupling term of the form $ R\phi ^2 $ is induced through quantum corrections. Recently, it was shown (see Ref. \cite{Elizalde:2015nya}) that starting from the action involving the term $ U(\phi ) R$, where $ U $ is a differentiable function of $ \phi $, the Einstein-Hilbert term can be generated through renormalization-group (RG) corrections.}
\begin{align}
	\label{specificCase}
	h(\phi )
	=
	\frac{\xi \phi ^2}{M_\text{Pl}^2}
	=
	\xi \phi ^2
	\quad
	\text{and}
	\quad
	V(\phi )
	=
	\frac{\lambda}{4}\phi ^4,	
\end{align}
corresponding to the nonminimal chaotic inflation model well-investigated in the references cited above and the references therein. This choice allows us to have a fixed model with well-known characteristics that we can exploit in our analysis. However, as it may be beneficial for future study of a more general case\footnote{A general treatment at this point could lead to questions about the feasibility of inflation,  behavior of the inflaton field, suitable energy range for inflation, among others, for all conceivable combinations of $ h(\phi ) $ and $ V(\phi ) $. These concerns that can be relevant in performing adiabatic regularization are beyond the scope of the current paper.}, we keep the symbol $ h(\phi ) $ and $ V(\phi ) $ in our calculations. The limit to the specific case given by (\ref{specificCase}) is considered in the final expression for the regularized power spectrum and in the discussion of some other quantities where there may be a need to have a fixed model.

For the Friedmann-Lema\^itre-Robertson-Walker (FLRW) metric describing the background spacetime for the action above\footnote{Needless to say, we are using the metric signature $ (-,+,+,+) $.}, 
\begin{align}
	\label{FLRWMetric}
	\dd s^2
	=
	-\dt^2
	+
	a^2(t)\delta _{ij}\dx^i\dx^j,
\end{align}
where $ a(t) $ is the scale factor, the equation of motion (EOM) for $ \phi $ reads
\begin{align}
	&\ddot{\phi } + 3H\dot \phi 
	+
	\bigg(
		1 + \frac{3}{2}\frac{h_\phi ^2}{f}
	\bigg)^{-1}V_\phi
	+
	\frac{1}{2}\frac{h_\phi }{f}(1 + 3h_{\phi \phi})\bigg(
		1 + \frac{3}{2}\frac{h_\phi ^2}{f}
	\bigg)^{-1}\dot \phi ^2
	\nonumber
	\\[0.5em]
	&\qquad
	-\,
	\frac{2h_\phi }{f}	\bigg(
		1 + \frac{3}{2}\frac{h_\phi ^2}{f} 
	\bigg)^{-1}V
	=
	0,
\end{align}
where the overdot means differentiation with respect to the coordinate time $ t $, and the subscript $ \phi $ denotes derivative with respect to $ \phi $. It is straightforward to verify that this equation reduces to the EOM for $ \phi  $ given in the paper of Komatsu and Futamase \cite{Komatsu:1999mt} for the case specified in (\ref{specificCase}). In addition to this, the Friedmann equation derived from the Einstein equation reads
\begin{align}
	H^2
	=
	\frac{\rho }{3}
	=
	\frac{1}{3f}\Big(
		\tfrac{1}{2}\,\dot \phi ^2
		+
		V
		-
		3\dot hH
	\Big),	
\end{align}
where $ H \equiv \dot a/a $, is the Hubble parameter.

The variation of the action with respect to $ g_{\mu \nu } $ leads to the expression for the conserved energy-momentum tensor $ T^\mu {}_\nu = \text{diag}(-\rho, p, p, p) $, with $ \rho  $ and $ p $ being the background energy density and pressure respectively.
\begin{align}
	\rho 
	&=
	\frac{1}{f}\Big(
		\tfrac{1}{2}\,\dot \phi ^2
		+
		V
		-
		3\dot hH
	\Big),
	\nonumber
	\\[0.5em]
	p
	&=
	\frac{1}{f}\Big(
		\tfrac{1}{2}\,\dot \phi ^2
		-
		V
		+
		2\dot h H
		+ 
		\ddot h			
	\Big).
\end{align}
Since $ \dot H = -\frac{1}{2}(\rho + p) $ and the slow-roll parameter $ \epsilon = -\dot H/H^2 $ by definition, we find
\begin{align}
	\label{hcontraint}
	\epsilon 
	=
	\frac{1}{f}\bigg(
		\frac{\dot \phi ^2}{2H^2}
		-
		\frac{\dot h}{2H}			
		+
		\frac{\ddot h}{2H^2}
	\bigg).
\end{align}
For the slow-roll (chaotic) inflation that we assume in this article, $ |\epsilon| \ll 1 $, implying the conditions $ \dot \phi ^2/fH^2,\,|\dot h/fH|,$ $|\ddot h/fH^2| \ll 1$. These inequalities are satisfied by the inflationary solution in the case where $ h(\phi ) = \xi \phi ^2 $ and $ V(\phi ) = (\lambda /4) \phi ^4$, with $ h \gg 1$. In particular, the Friedmann equation reads \cite{Fakir:1990eg,Komatsu:1999mt}, 
\begin{align}
	H^2
	=
	\frac{\lambda h}{12\xi ^2},
	\quad
	\text{with}
	\quad
	\frac{\dot h}{H}
	=
	-\frac{8\xi }{1 + 6\xi },
\end{align}
as a consequence of standard slow-roll assumption, the condition $ h \gg 1 $, and the equation of motion for $ \phi $. Inflation ends when $ \epsilon = 1 $ corresponding to $ h = \xi \phi ^2 \sim \mathcal O(1) $ \cite{Fakir:1990eg}.

\bigskip 
\bigskip
\section{The ``bare'' Power Spectrum}
\label{``bare''Power}
With the background equations in place, we now consider the perturbations and proceed to calculate their power spectra. In our calculations, we use the \textit{comoving gauge} where the inflaton fluctuation vanishes ($ \delta \phi = 0 $) and the three-dimensional metric takes the form
\begin{align}
	g_{ij}
	&=
	a^2(t)e^{2\mathcal R(\mathbf x, t)}\left(e^{\gamma(\mathbf x, t)} \right)_{ij},
	\quad
	\partial ^i\gamma _{ij} = \gamma ^i{}_{i} = 0,
\end{align}
where $ \mathcal R $ is the gauge-invariant (scalar) curvature perturbation and $ \gamma _{ij} $ is the transverse traceless tensor perturbation. Including these perturbations about the homogeneous and isotropic spacetime described by the FLRW metric (\ref{FLRWMetric}), the metric in the Arnowitt-Deser-Misner (ADM) formulation (that we use here,) can be written as \cite{Arnowitt:1959ah,Gourgoulhon}
\begin{align}
	\dd s^2
	&=
	-N^2\dt^2
	+
	g_{ij}(\dx^i + N^i\dt)(\dx^j + N^j \dt),
\end{align}
where $ N $ and $ N^i $ are the lapse and shift functions respectively.

Using the metric above and the comoving gauge, the computation of the power spectrum\footnote{The calculation for the tensor perturbation follows effectively the same route.} for $ \mathcal R $ proceeds by the decomposition of the action as $ S = S^{(0)} + S^{(2)}  + S^{(3)} + \cdots, $ where the quantity $ \textit{n} $ in $ S^{(n)} $ indicates the order with respect to $ \mathcal R $ (see for instance, Refs. \cite{Chen:2010xka,Qiu:2011}). The segment $ S^{(2)} $ corresponds to the power spectrum. In the minimal case, the EOM called the\textit{ Mukhanov-Sasaki} equation, derived from $ \bar S^{(2)} $ reads
\begin{align}
	\label{MukhanovSasakiEqn}
	\bar v_k''
	+
	\left(
		k^2
		-
		\frac{\bar z''}{\bar z}
	\right)\bar v_k
	=
	0,
\end{align}
where the overbar signifies minimal coupling, the subscript $ k $ indicates momentum space, the prime symbol means differentiation with respect to the conformal time ($ \dd\tau \equiv \dt/a $), $ \bar{\mathcal R} = \bar v/\bar z $ and $ \bar z^2 \equiv 2a^2\bar \epsilon$. Unfortunately, the decomposition in the nonminimal coupling case is complicated by the presence of the additional term $ h(\phi )R $ in the action leading to the more challenging derivation of the Mukhanov-Sasaki equation. Such a complication is remedied by transforming from the current frame of calculation called the \textit{Jordan frame}, to the \textit{Einstein frame}.

In the Einstein frame, the action takes the form of the minimal case resulting in greatly simplified calculation. The two frames are related through the metric conformal transformation $ \widehat{g}_{\mu \nu } = \Omega ^2 g_{\mu \nu } $, where we take $ \Omega ^2 = f(\phi ) = 1 + h(\phi ) $ and the hat means a given quantity is evaluated in the Einstein frame. As a consequence, 
\begin{align}
	\label{JordanEinstein}
	\widehat a 
	&= 
	\Omega a,
	\quad
	\dd\widehat t
	=
	\Omega \dt,
\end{align}
and we may define quantities in analogy with those defined in the Jordan frame; e.g.,
\begin{align}
	\label{JordanEinstein2}
	\widehat H
	=
	\frac{1}{\widehat a}
	\frac{\dd \widehat a}{\dd \widehat t},
	\quad
	\widehat \epsilon 
	=
	-\frac{1}{\widehat H^2}
	\frac{\dd\widehat H}{\dd\widehat t}.
\end{align}
The primordial cosmological perturbations are fortunately, frame invariant; that is, $ \widehat {\mathcal R} =	\mathcal R $ \cite{Makino:1991sg,Fakir:1990eg,Kubota}. Furthermore, the decomposition of the action in the Jordan frame is correspondingly reflected in the Einstein frame as $ \widehat S = \widehat S^{(0)} + \widehat S^{(2)}  + \widehat S^{(3)} + \cdots, $ with $ \widehat S^{(n)} = S^{(n)} $ \cite{Kubota}. Consequently, the Mukhanov-Sasaki equation can be written as
\begin{align}
	\label{MSEqEinstein}
	\widehat  v_k''
	+
	\left(
		k^2
		-
		\frac{\widehat  z''}{\widehat  z}
	\right)\widehat v_k
	=
	0,
	\qquad
	\big(\widehat z^2 = 2\widehat{a}^2\widehat \epsilon\big) 
\end{align}
from which one can derive the expression for the curvature perturbation,
\begin{align}
	{\mathcal R_k}
	=
	\frac{\widehat v_k}{\widehat z},
\end{align}
which in turn, gives us the ``bare'' power spectrum:
\begin{align}
	\label{powSpec``bare''}
	\Delta^{2(b)} _{\mathcal R}
	=
	\frac{k^3}{2\pi ^2}\big|\mathcal R_k\big|^2\bigg|_{k = aH}.
\end{align}

It then remains for us to express the potential $ \widehat z''/\widehat z $ in terms of the conformal time and the corrections due to the slow-roll parameters and the nonminimal coupling terms, to solve for $ \widehat v_k $ which will give us $ \mathcal R_k $. In our previous work on adiabatic regularization in minimally coupled general single-field inflation \cite{Alinea:2015pza}, we used the result in Ref. \cite{Zhu:2014wfa}\footnote{As of the time of writing, this is the most precise calculation based on uniform approximation extending the computation in Ref. \cite{Martin:2013uma} for $ k $-inflation.} based on the promising semi-analytical method of solving the Mukhanov-Sasaki equation called the \textit{uniform approximation} \cite{Habib:2002yi,Habib:2004kc}. But as we showed in Ref. \cite{Alinea:2015gpa}, the method as implemented in Refs. \cite{Martin:2013uma,Zhu:2014wfa} utilizing a certain expansion scheme for the slow-roll parameters, for the minimally coupled general single-field inflation, can lead to logarithmic divergences in the expression for the power spectrum. As such, as long as the problem with the implementation of the uniform approximation in relation with the use of suitable expansion scheme for the slow-roll parameters is not properly addressed, other semi-analytic approaches such as Green's function method \cite{Gong:2001he} and WKB approximation \cite{Martin:2002vn,Casadio:2005em} may be preferred. In this work however, we follow the more modest and common approach using the Hankel function approximation in line with Ref. \cite{Qiu:2011} and in combination with the method of frame transformation. For the most part, if only to see the behavior of the regularised power spectrum, we would not be needing an extremely precise expression for the ``bare'' part and the subtraction term.

To proceed with the calculation of the potential $ \widehat z''/\widehat z $, by virtue of (\ref{JordanEinstein}), (\ref{JordanEinstein2}), and the relation $ \Omega ^2 = 1 + h $, we express $ \widehat \epsilon  $ as
\begin{align}
	\label{epsilonHat}
	\widehat \epsilon 
	&=
	\frac{\bar \epsilon }{\Omega ^2}\bigg(
		1 + \frac{\dot h}{2H\Omega ^2}
	\bigg)^{-2}\bigg(
		1
		+
		\frac{3}{2}\frac{h_\phi ^2}{\Omega ^2}
	\bigg)
	=
	\frac{\bar \epsilon }{\Omega ^2}\frac{p_2}{p_1^2},
\end{align}
where $ \bar \epsilon $ is the familiar slow-roll parameter in the minimal case ($ h = 0 $). For brevity, we have denoted
\begin{align}
	\label{p1p2}
	p_1
	\equiv
	1 + \beta 
	\quad
	\text{and}
	\quad
	p_2
	\equiv
	\bigg(
		1 + \frac{3}{2}\frac{h_\phi ^2}{\Omega ^2}
	\bigg),
\end{align}
with the other slow-roll parameter $ \beta \equiv \dot h/2H\Omega ^2$. (This is the same slow-roll parameter $ \beta $ introduced in Ref. \cite{Komatsu:1999mt}.) The quantity $ \widehat z $ can then be written as
\begin{align}
	\label{zHat}
	\widehat z 
	&=
	\frac{a\sqrt{2\bar \epsilon p_2}}{p_1},
\end{align}
from which follows
\begin{align}
	\label{potentialZ}
	\frac{\widehat z\,''}{\widehat z}
	&=
	(aH)^2\bigg[
		2
		-
		\epsilon 
		+
		\frac{3}{2}\frac{\dot{\bar \epsilon }}{H\bar \epsilon }
		+
		\frac{\ddot{\bar \epsilon }}{2H^2 \bar \epsilon }
		-
		\frac{\dot{\bar \epsilon }^2}{4H^2\bar \epsilon ^2}
		-
		\bigg(
			\frac{3}{H}
			+
			\frac{\dot{\bar \epsilon }}{H^2\bar \epsilon }
		\bigg)\bigg(
			\frac{\dot p_1}{p_1}
			-
			\frac{\dot p_2}{2p_2}
		\bigg)
		\nonumber
		\\[0.5em]
		&\qquad\qquad
		-\,
		\frac{1}{H^2}\bigg(
			\frac{\ddot p_1}{p_1}
			-
			\frac{\ddot p_2}{2p_2}
		\bigg)
		+
		\frac{1}{H^2}\bigg(
			\frac{2\dot p_1^2}{p_1^2}
			-
			\frac{\dot p_1\dot p_2}{p_1p_2}
			-
			\frac{\dot p_2^2}{4p_2^2}
		\bigg)
	\bigg],
	\nonumber
	\\[0.5em]
	\frac{\widehat z\,''}{\widehat z}
	&=
	(aH)^2\Big[2 - \epsilon + \tfrac{3}{2}\bar \epsilon _2 
	+ 
	\epsilon _s 
	+
	\mathcal O(\epsilon ^2)
	\Big],
\end{align}
where $ \epsilon _s = \epsilon _s(p_1, p_2, \dot p_1, \dot p_2)$ captures the first order terms due to the nonminimal coupling term, and $ \bar \epsilon _2 $ is defined as $ \dot {\bar \epsilon} = H\bar \epsilon \bar \epsilon _2 $. The slow-roll parameter $ \bar \epsilon _2 $ is a part of the chain of Hubble flow parameters, $ \dot {\bar\epsilon} _n \equiv H\bar \epsilon _n\bar \epsilon _{n + 1},\, n = 1, 2, 3, \cdots $, introduced in Ref. \cite{Schwarz:2001vv}.

Since the conformal time can be expressed in terms of $ (aH) $ and the slow-roll parameter(s) as \cite{Alinea:2015gpa}
\begin{align}
	\tau 
	&=
	-\frac{1}{a}\sum_{n=0}^\infty\left(
		H^{-1}\frac{\dd}{\dt}
	\right)^n H^{-1}
	=
	-\frac{1}{aH}(1 + \epsilon + \cdots ),
\end{align}
then together with the last of (\ref{potentialZ}), we can rewrite the Mukhanov-Sasaki equation in a familiar form as
\begin{align}
	\label{MSv}
	\widehat  v_k''
	+
	\left[
		k^2
		-
		\frac{1}{\tau ^2}\Big(
			2 
			+ 
			3\tilde \epsilon_s 
			+
			\tfrac{3}{2}\,\bar \epsilon _2
		\Big)
	\right]\widehat v_k
	=
	0,
\end{align}
where $ \tilde \epsilon_s \equiv \epsilon + \epsilon _s/3  $. The solution of this differential equation can be written in terms of the Hankel functions the limiting form of which for the superhorizon modes ($ k\ll aH $) can be expressed in terms of the gamma function. We find
\begin{align}
	\widehat v_k
	\simeq
	e^{i(2\nu + 3)\frac{\pi }{4}}
	\frac{2^{\nu -\frac{3}{2}}\Gamma (\nu )}{\sqrt{2k}\Gamma (\frac{3}{2})}
	(1 - \epsilon )^{\nu - \frac{1}{2}},
\end{align}
where $ \nu \equiv \tfrac{3}{2} + \tilde \epsilon_s + \tfrac{1}{2}\bar \epsilon _2$.

Expanding the gamma function above and the quantities involving powers of $ \nu  $, we find the square of the modulus of $ \mathcal R_k $ as
\begin{align}
	|\mathcal R_{k*}|^2
	\simeq
	\frac{1}{4ka_*^2p_{2*}\bar \epsilon _{*}}(1 + \delta \epsilon _{s*}),
\end{align}
where the symbol `*' signifies evaluation at the horizon crossing and $	\delta \epsilon _{s*} =	(2\tilde \epsilon_{s*} + \bar \epsilon_{2*} )(2 - \gamma - \ln 2) - 2\epsilon_* + 2\beta _*$, with $ \gamma = 0.5772 $ being the Euler-Mascheroni constant. The ``bare'' power spectrum is now straightforward to calculate using these results and the definition (\ref{powSpec``bare''}). We have\footnote{The result below can be also obtained by frame transformation from the Einstein frame utilizing the minimal result, to the Jordan frame. Following this path however, one has to be careful with the relations between the Hubble flow parameters in the Einstein and Jordan frames. Our calculation in this paper including that for the subtraction term, follows a self-contained derivation.}
\begin{align}
	\label{``bare''PowSpec}
	\Delta^{2(b)} _{\mathcal R}
	&=
	\frac{H_*^2}{8\pi ^2p_{2*}\bar \epsilon _*}(1 + \delta \epsilon _{s*}).
\end{align}
For $ h(\phi ) = 1 + \xi \phi ^2 $, this becomes
\begin{align}
	\Delta^{2(b)} _{\mathcal R}
	&=
	\frac{H_*^2}{8\pi ^2(1 + 6\xi )\bar \epsilon _{*}}
	(1 + \delta \epsilon _{s*}).
\end{align}
The extra factor $ (1 + 6\xi) $ (that seems to have been missed or intentionally omitted in Ref. \cite{Nozari:2010uu}) signifies the generalization to the nonminimal coupling case. Apart from the correction $ \delta \epsilon _{s*} $, it is exactly the same as that obtained by Makino and Sasaki \cite{Makino:1991sg} and Komatsu and Futamase \cite{Komatsu:1999mt} through a slightly different method. The more precise calculation in the Jordan frame utilising the Hankel function approximation can be found in Ref. \cite{Qiu:2011}. But as the final expression for the power spectrum involves the evaluation of $ \Omega $ at an unusual value of $ \tau = -1 $ and different parameters, it may not be practical to do the comparison\footnote{Moreover, some terms first order in $ \epsilon  $ seem to have been (perhaps unintentionally) neglected.}.

The calculation for the ``bare'' tensor power spectrum follows the same path as that of the scalar perturbation. We have
\begin{align}
	\Delta _\gamma ^{2(b)}
	&=
	\frac{k^3}{\pi ^2}|h_k|^2\bigg|_{k = aH},
\end{align}
where $ h_k $ is the amplitude of tensor perturbation. The EOM valid up to first order in $ \epsilon  $, in the Einstein frame\footnote{The second-order action in the Jordan frame in the more general setting where there is a nonminimal derivative coupling, is given in Ref. \cite{Nozari:2015lia}.}, reads
\begin{align}
	\widehat \mu _k''
	+
	\left[
		k^2 - \frac{1}{\tau ^2}(2 + 3\tilde \epsilon_t )
	\right]\widehat \mu _k
	&=
	0,	
\end{align}
with $ \widehat h_k = \widehat \mu _k/\widehat a,\,\tilde \epsilon_t = \epsilon + \epsilon _t/3 $, and 
\begin{align}
	\epsilon_t
	=
	\frac{3\dot h}{2H\Omega ^2}
	+
	\frac{\ddot h}{2H^2\Omega ^2}
	-
	\frac{\dot h^2}{4H^2\Omega ^4}\,.
\end{align}
Solving the differential equation then leads to the sought-for expression for the ``bare'' tensor power spectrum given by
\begin{align}
	\Delta _\gamma ^{2(b)}
	=
	\frac{2H_*^2}{\pi^2\Omega ^2_*}(1 + \delta \epsilon _{t*}),
\end{align}
where $ \delta \epsilon _{t*} = 2\tilde \epsilon _{t*}(2 - \gamma - \ln 2) - 2\epsilon_* $. Note the residual term $ \Omega _* $ in the denominator not present (in the same position) in the power spectrum for the scalar perturbation. Such a term arises from the absence of $ \widehat \epsilon  $ in the amplitude $ \widehat h_k = \widehat \mu _k/\widehat a$ for the tensor perturbation. This is in contrast with that of the scalar perturbation where the amplitude of the perturbation is $ \widehat v_k/(\widehat a \sqrt{2\widehat \epsilon }\,)$; the quantity $ \Omega $ in $ \widehat a $ exactly cancels that in $ \sqrt{\widehat \epsilon } $.

\bigskip
\bigskip
\section{Adiabatic Regularization of the Power Spectra}
\label{adiabaticReg}
In this section we wish to derive the subtraction term and examine the behavior of the regularized power spectrum. Similar to that of the ``bare'' power spectrum, such a derivation is complicated by the existence of the nonminimal coupling term $ h(\phi )R $ in the action. We have dealt with such a complication above by using the frame-invariance of $ \mathcal R $. It would then be good if such a simple relationship would also hold for the subtraction term. Fortunately, as what we have shown in our previous work \cite{Alinea:2015pza}, in adiabatic regularization, this is indeed the case; i.e., $ \widehat {\mathcal R}^{(s)} = \mathcal R^{(s)} $. Needless to say, the same holds for the tensor perturbation.

In performing adiabatic regularization (see Refs. \cite{Parker:Toms,BirrellDavies} for the details of the steps that we simply outline here), one may start with the ansatz
\begin{align}
	v_k^{(s)}(\tau )
	=
	\frac{1}{\sqrt{2W}}
	e^{-i\int ^{\tau }\dd\tau' W(\tau ')},
\end{align}
whose form is consistent with the short-wavelength limit of $ v_k^{(b)} (\tau )$, where $ \mathcal R^{(b)} = v_k^{(b)}/z $ and $ \mathcal R^{(s)} = v_k^{(s)}/z $, guaranteed by the Mukhanov-Sasaki equation given by (\ref{MSEqEinstein}). The differential equation satisfied by $ v_k^{(s)}(\tau ) $ used in conjunction with the Mukhanov-Sasaki equation gives the working equation needed to solve for $ W $ to a certain desired adiabatic order. This then allows us to solve for $ \mathcal R^{(s)} $ up to an adiabatic order consistent with the prescription of the minimal subtraction scheme; considering the UV divergence(s) in the two-point function of $ \mathcal R(\tau, \mathbf x) $, cf. (\ref{twoPtFn}), for the power spectrum, it is up to second adiabatic order. 

The working equation for $ W $ in the minimal-coupling case reads
\begin{align}
	W^2
	-
	\frac{3}{4}\left(
		\frac{W'}{W}
	\right)^2
	+
	\frac{1}{2}\frac{W''}{W}
	-
	\left(
		k^2
		-
		\frac{\bar z''}{\bar z}
	\right)
	=
	0.
\end{align}
To find $ W $, one expands it as $ W = \omega _0 + \omega _1 + \omega _2 + \cdots, $ where $ n $ in $ \omega _n $ denotes the adiabatic order, and solve the resulting differential equation order-by-order. As it turns out,
\begin{align}
	\omega _0
	&=
	k,
	\quad
	\omega _1
	=
	0,
	\quad
	\omega _2
	=
	-\frac{1}{2k}\frac{\bar z''}{\bar z}.
\end{align}
To deal with the nonminimal-coupling case, we simply have to replace $ \bar z $ by $ \widehat z $. Upon performing substitution from the equation for $ W $ to the equation for $ v_k^{(s)} $ and then to the expression for $ \widehat {\mathcal R}_k $, we find 
\begin{align}
	\big|{\mathcal R}^{(s)}_k\big|^2
	&=
	\big|\widehat {\mathcal R}^{(s)}_k\big|^2
	=
	\frac{1}{2k\widehat z^2}\left(
		1 + \frac{1}{2k^2}\frac{\widehat z''}{\widehat z}
	\right).
\end{align}

The expressions for the quantity $ \widehat z $ and the potential $ \widehat z''/\widehat z $ are given by (\ref{zHat}) and the second of (\ref{potentialZ}) respectively. Substituting these relations in the last equation above and noting that we are dealing with superhorizon modes ($ k \ll aH $), we find the subtraction term (after the multiplication by $ k^3/2\pi ^2 $) as
\begin{align}
	\label{subTerm}
	\Delta ^{2(s)}_{\mathcal R}
	=
	\frac{H^2}{8\pi ^2p_2\bar\epsilon }(1 + \delta \epsilon _s),
\end{align}
where $	\delta \epsilon _{s} =	2\beta -\frac{1}{2}\epsilon + \frac{3}{4}\bar \epsilon _2 + \frac{1}{2}\epsilon _s $. This expression takes the same form as that of the ``bare'' power spectrum given by (\ref{``bare''PowSpec}). What distinguishes these two expressions from the minimal case is the presence of $ p_2 $ and the more general expression for $ \delta \epsilon $ that includes $ \Omega $ and its derivative. 

The difference between the ``bare'' power spectrum and the subtraction term gives the regularized power spectrum, $ \Delta ^{2(r)}_{\mathcal R} $. We have from (\ref{``bare''PowSpec}) and (\ref{subTerm}),
\begin{align}
	\Delta ^{2(r)}_{\mathcal R}
	&=
	\frac{H_*^2}{8\pi ^2p_{2*}\bar \epsilon _*}\left[
		(1 + \delta \epsilon _{s*})
		-
		(1 + \delta \epsilon _s)
		\frac{H^2}{H_*^2}
		\frac{\bar \epsilon_*}{\bar \epsilon}
		\frac{p_{2*}}{p_2}
	\right].
\end{align}
In the canonical case, $ p_2 $ is unity and $ \delta \epsilon \sim \mathcal O(\bar \epsilon )$; the equation above reduces to the result derived by Urakawa and Starobinsky \cite{Urakawa:2009}. In the nonminimal case in hand, $ |\delta \epsilon| \sim \mathcal O(\epsilon ) $ is likewise small so we simply have to mind the behavior of $ p_2,\,H,\,\text{and}\, \bar \epsilon$. For $ h = \xi \phi ^2 $, inflation starts at some relatively large value of $ \xi \phi ^2 $ and ends when $ \xi \phi ^2 \sim \mathcal O(1) $ \cite{Fakir:1990eg}. It follows that (see the second of (\ref{p1p2}))
\begin{align}
	p_2
	=
	\frac{1 + \xi \phi ^2 + 6\xi ^2\phi ^2}{1 + \xi \phi ^2}
	\quad
	\rightarrow
	\quad
	\sim\,
	1 + 3\xi,
\end{align}
while $ p_{2*} \simeq 1 + 6\xi $. However, the term involving the Hubble parameter behaves as
\begin{align}
	\label{behaveH}
	\frac{H^2}{H_*^2}
	&=
	e^{-2\int_{N*}^N \dd N'\epsilon (N)},
\end{align}
where $ N $ is the number of $ e $-folds with $ \dd N \equiv \dd \ln a$.
The slow-roll parameter $ \epsilon $ goes from near zero at the start of inflation, to unity at the end of inflation, resulting in the decay of the factor $ H^2/H_*^2 $. Likewise, since $ \bar \epsilon $ increases (at least on average,) as inflation progresses, the factor $ \bar \epsilon_*/\bar \epsilon $ decreases with time. Hence, if we follow the the exponentially decaying subtraction term long enough, the regularized power spectrum becomes simply the ``bare'' power spectrum. This is the same conclusion found in the minimal case \cite{Urakawa:2009,Alinea:2015pza}. The presence of the nonminimal coupling term in the form of $ \xi \phi ^2R $ does not significantly alter the exponentially decaying behavior of the subtraction term and has no significant effect on the regularized power spectrum of the scalar perturbation.

For tensor perturbations, following the same procedure as that of the scalar perturbation, we find the regularized power spectrum as
\begin{align}
	\Delta ^{2(r)}_\gamma 
	&=
	\frac{2H_*^2}{\pi ^2\Omega _*^2}\left[
		1 + \delta \epsilon _{t*}
		-
		(1 + \delta \epsilon _t)
		\frac{H^2}{H_*^2}
		\frac{\Omega_* ^2}{\Omega ^2}
	\right],
\end{align}
where $ \delta \epsilon _{t} = -\frac{1}{2}\epsilon + \frac{3}{2}\beta  $. The conformal factor is $ \Omega ^2 = 1 + \xi \phi ^2 $. During inflation, $ \xi \phi ^2 $ decreases so the factor $ \Omega _*^2/\Omega ^2 $ increases. However, as is apparent from the previous paragraph, $ \Omega ^2 $ is bounded from below with $ \xi \phi ^2 \sim \mathcal O(1)$ marking the end of inflation. Furthermore, $ \xi \phi ^2 \sim N_t $, where $ N_t $ is the total number of $ e $-folds at ``the time where the scales of astronomical relevance cross outside the Hubble radius'' \cite{Fakir:1990eg}. This gives us an estimate of $ \Omega _*^2/\Omega ^2 \sim \mathcal O(N_t)$ long after horizon crossing. It follows that the decreasing term $ \Omega _*^2/\Omega ^2 $ cannot compete with the exponentially decaying factor $ H^2/H_*^2 $ in the long run. Hence, as is the case for scalar perturbation, if we follow the subtraction term long enough, the regularized power spectrum tends to the ``bare'' power spectrum. 

\bigskip
\bigskip
\section{Remarks on the Use of the Einstein Frame}
\label{remarksEinsteinFrame}
\begin{figure}[hbt]
	\centering
	\makebox[\textwidth][c]{\includegraphics[scale=0.85]{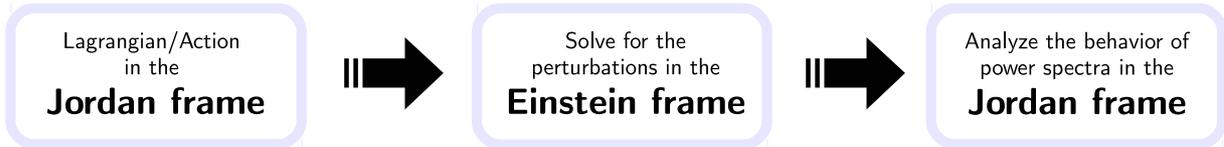}}	
	\caption{Flow diagram representing the procedure of regularization we followed in this work in relation to the use of the Einstein and Jordan frames.}
	\label{figFlowOfReg}
	\bigskip
\end{figure}

The entire regularization procedure we have followed in this work involves a flow from the Jordan frame to the Einstein frame and then back to the Jordan frame (see \textbf{Fig}. \ref{figFlowOfReg}). We started with the Lagrangian for nonminimally coupled inflation given by (\ref{lagrangianNonminimal}). Needless to say, this is within the framework of the Jordan frame where inflation means $ \ddot a > 0 $ and $ \epsilon $ goes from near zero to unity marking the end of inflation. Then, owing to the difficulty of solving for the perturbations in the Jordan frame, we calculated them in the Einstein frame and exploited the fact that $ \mathcal R_k $ and $ \gamma ^{ij}_k $ are the same in both frames. Finally, we performed the subtraction process for the resulting power spectra (``bare'' minus the subtraction term), taking into account the behavior of the subtraction term, in the Jordan frame.

In doing the final part, we find it extremely helpful if not crucial to do the subtraction process in the Jordan frame. In performing adiabatic regularization of the power spectrum, the knowledge of the behavior of several important parameters ($ H, \epsilon  $, etc.) from horizon crossing until at least the end of inflation is necessary. The information about these come in handy in the original frame. We know (or at least we can calculate them) that $ \epsilon  \sim 0 \rightarrow 1$ during inflation and so are the behavior of $ H $ and other parameters from the time corresponding to $ N_* $, onwards. 

Now suppose we consider the regularization in the Einstein frame. In order to perform the regularization in this frame, we should properly transform the parameters involved in our original model to the Einstein frame. This however, poses some problems. Based on (\ref{JordanEinstein}), $ \dd^2 a/\dd t^2 > 0$ does not necessarily imply $ \dd^2 \widehat a/\dd\widehat t\,^2 > 0$; that is, inflation in the Jordan frame does not necessarily imply inflation in the Einstein frame. Furthermore, by virtue of (\ref{JordanEinstein2}),
\begin{align}
	\widehat \epsilon 
	&=
	\bigg(
		1 + \frac{1}{H}\frac{\dot \Omega }{\Omega }
	\bigg)^{-2}\bigg[
		\epsilon 
		+
		\frac{1}{H}\frac{\dot \Omega }{\Omega }
		+
		\frac{1}{H^2}\bigg(
			2\frac{\dot \Omega ^2}{\Omega ^2}
			-
			\frac{\ddot\Omega }{\Omega}
		\bigg)
	\bigg],
\end{align}
we know that $ \epsilon  \sim 0 \rightarrow 1$ does not necessarily imply $ \widehat \epsilon  \sim 0 \rightarrow 1$, nor is the average monotonically increasing behavior of $ \epsilon $ crucial in establishing the exponentially decaying behavior of the Hubble parameter (see the previous section).\footnote{Interested readers may want to consult Ref. \cite{Nozari:2010uu} for a more thorough treatment of inflationary parameters and their differences in the Einstein and Jordan frames.} While we know the behavior of the inflationary parameters in the Jordan frame, the same is not straightforward when we consider the transformed parameters ($ H\rightarrow \widehat H, \epsilon\rightarrow \widehat \epsilon, N_* \rightarrow \widehat N_* $, etc.) in the Einstein frame. This is further complicated by the difference in the times of horizon crossing in the two frames. It is thus, far more computationally economical to do the final subtraction process in the Jordan frame where we have the original model in place.

The non-trivial mapping of inflationary parameters from the Jordan frame to the Einstein frame also tells us one important point: if the subtraction term in the regularization of the power spectrum tends to zero in the minimal case, this does not necessarily mean that the same holds in the nonminimal case. Although the result of Urakawa and Starobinsky \cite{Urakawa:2009} tells us that the regularization of the power spectrum in the canonical inflation yields the ``bare'' power spectrum, this does not necessarily mean that all corresponding nonminimally coupled inflation transformable to the canonical case through frame transformation should yield the same result. Whereas we take $ \epsilon \sim 0 \rightarrow \sim 1$ in the canonical case for the subtraction term, we take the \textit{same} ($ \epsilon \sim 0 \rightarrow \sim 1$) for the nonminimal case; yet in going from the Jordan frame to the Einstein frame for the nonminimal case, we cannot simply take $ \epsilon \sim 0 \rightarrow \sim 1$ to $ \widehat \epsilon \sim 0 \rightarrow 1 $. This is not to say that the adiabatic regularization in the Jordan and Einstein frames would yield different results.\footnote{See Refs. \cite{Faraoni:2006fx,Nozari:2010uu,Faraoni:1999hp} for a slightly different perspective regarding the equivalence of the Einstein and Jordan frames at the quantum level. The authors suggest that at the quantum level, there can be differences between the two frames.} But to emphasize that at least within the framework of adiabatic regularization, the canonical case studied by Urakawa and Starobinsky \cite{Urakawa:2009} does \textit{not exactly} correspond to the Einstein frame of the nonminimal case in this study; the behavior of the inflationary parameters may differ significantly.

With this being said, the foregoing seems to only scratch the surface of subtleties associated with frame transformation from the perspective of adiabatic regularization. It would be interesting to further explore the behavior of the subtraction term in the Einstein frame and how it relates to that in the Jordan frame. But as the nontrivial mapping between the inflationary parameters suggests computational and conceptual complexities requiring more space than we can allot here, we wish not to deviate from the main concern of this work. Looking ahead, this will be another direction along which we can extend this study.

\bigskip
\bigskip
\section{Concluding Remarks}
\label{conclude}
Power spectrum is one of the most important quantities in inflationary cosmology. In this work, we investigated the effect of ultraviolet (UV) regularization on the power spectrum within the framework of nonminimally coupled chaotic inflation. Our calculation suggests that the adiabatic regularization leads to no difference between the ``bare'' and regularized power spectra for both scalar and tensor perturbations. Going beyond the time of horizon crossing, the subtraction term exponentially decays with respect to the number of $ e $-folds and becomes insignificant in the long run; the expansion of the Universe essentially erases the subtraction term. This is the same result we obtained for minimally coupled general single-field inflation in our previous study \cite{Alinea:2015pza}. Thus, for both the minimally coupled and nonminimally coupled cases, the current and the previous studies together with that of Urakawa and Starobinsky \cite{Urakawa:2009}, justify the use of the ``bare'' power spectrum in standard calculations.

So far, we have only investigated the area of nonminimally coupled inflation where the speed of sound is constant. It would be interesting to see the effect of varying speed of sound $ c_s^2 $, in scenarios where the usual kinetic-plus-potential term is replaced by a general ``pressure'' function $ P(X, \phi ) $, where $ X \equiv -\frac{1}{2}g^{\mu \nu }\phi _{,\mu }\phi _{,\nu } $. The subtraction term would significantly be more complicated because of the involvement of the time derivatives of $ c_s^2 $ and the more complex relationship between the speed of sound in the Jordan and Einstein frames. From a bird's eye view, three main terms have to be considered namely, those due to (a) the speed of sound, (b) variation of the Hubble parameter, and (c) the effects of the nonminimal coupling term. Beyond this scenario, one may also consider nonminimal inflation beyond Einstein gravity; e.g., instead of $ h(\phi )R $, one may consider $ h(\phi )f(R) $\footnote{Such generalization may be relevant when quantum corrections are considered. For instance, in Ref. \cite{Salvio:2015kka}, it is shown that at the quantum level, ``the renormalization for large non-minimal coupling requires an additional degree of freedom that transforms Higgs inflation into Starobinsky $ R^2 $ inflation, unless a tuning of the initial values of the running parameters is made.''}, where $ f(R) $ is a function of the Ricci scalar. In addition to this, we have not addressed the more challenging problems involving interactions and loop corrections. Considering all these problems, we look ahead to establishing a general set of conditions within the framework of general single-field inflation, both minimal and nonminimal, that would guarantee the null effect of adiabatic regularisation on both scalar and tensor power spectra.

\acknowledgments
Much of the core ideas in this work developed during the early stages of research involving minimally coupled general single-field inflation wherein the author collaborated with Dr. Wade Naylor, Yukari Nakanishi, and Prof. Takahiro Kubota of High Energy Theory laboratory, Osaka University (Japan). The author would like to particularly express his gratitude to Prof. Kubota for lending his ear, constructive comments about this work, and mentorship in general. In addition to this, the author would like to thank Dr. S. Vernov and Dr. E. Pozdeeva of Lomonosov Moscow State University for pointing out some typographical errors in the earlier version of the manuscript.


\end{document}